\documentclass[5p,twocolumn]{elsarticle}

\usepackage{amssymb}

\journal{Nuclear Instruments and Methods A}

\begin{document}

\begin{frontmatter}



\title{Fabrication of silica aerogel with $n$ = 1.08 for $e^+/\mu ^+$ separation in a threshold Cherenkov counter of the J-PARC TREK/E36 experiment}


\author[First]{Makoto Tabata\corref{cor1}}
\ead{makoto@hepburn.s.chiba-u.ac.jp}
\cortext[cor1]{Corresponding author.} 
\author[Second]{Akihisa Toyoda}
\author[First]{Hideyuki Kawai}
\author[Second]{Youichi Igarashi}
\author[Second]{\\Jun Imazato}
\author[Third]{Suguru Shimizu}
\author[Fourth]{Hirohito Yamazaki\fnref{fn1}}
\fntext[fn1]{Present address: Radiation Science Center, High Energy Accelerator Research Organization (KEK), Tsukuba, Japan}

\address[First]{Department of Physics, Chiba University, Chiba, Japan}
\address[Second]{Institute of Particle and Nuclear Studies (IPNS), High Energy Accelerator Research Organization (KEK), Tsukuba, Japan}
\address[Third]{Department of Physics, Osaka University, Toyonaka, Japan}
\address[Fourth]{Research Center for Electron Photon Science, Tohoku University, Sendai, Japan}

\begin{abstract}
This study presents the development of hydrophobic silica aerogel for use as a radiator in threshold-type Cherenkov counters. These counters are to be used for separating positrons and positive muons produced by kaon decay in the J-PARC TREK/E36 experiment. We chose to employ aerogel with a refractive index of 1.08 to identify charged particles with momenta of approximately 240 MeV/$c$, and the radiator block shape was designed with a trapezoidal cross-section to fit the barrel region surrounding the kaon stopping target in the center of the TREK/E36 detector system. Including spares, we obtained 30 crack-free aerogel blocks segmented into two layers, each layer having a thickness of 2 cm and a length of 18 cm, to fill 12 counter modules. Optical measurements showed that the produced aerogel tiles had the required refractive indices and transparency.
\end{abstract}

\begin{keyword}
Silica aerogel \sep Cherenkov radiator \sep Refractive index \sep Particle identification \sep J-PARC TREK/E36

\end{keyword}

\end{frontmatter}



\section{Introduction}
\label{}
\begin{figure}[h]
\centering 
\includegraphics[width=0.40\textwidth,keepaspectratio]{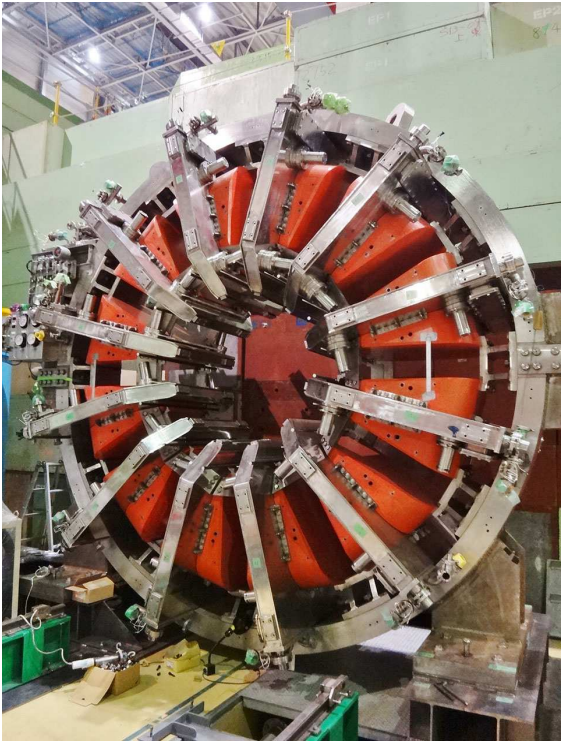}
\caption{A 12-sector superconducting iron-core toroidal spectrometer installed in the K1.1BR beam line of J-PARC in November 2014. This spectrometer has 12 identical gaps and a rotational symmetry of 30$^\circ $.}
\label{fig:fig1}
\end{figure}

For the purposes of the J-PARC E36 experiment \cite{cite1} by the TREK Collaboration \cite{cite2}, we are upgrading the experimental E246 apparatus \cite{cite3,cite4}---which was based on a 12-sector superconducting iron-core toroidal spectrometer (Fig. \ref{fig:fig1}) \cite{cite5} previously used at the High Energy Accelerator Research Organization (KEK), of Tsukuba, Japan---into a new TREK/E36 detector system. The primary goal of the TREK/E36 experiment is to test lepton universality using the decay channel $K^+ \to l^+\nu_l$ (known as $K_{l2}$ decay), where $l = e$ or $\mu $. To search for new physics beyond the Standard Model, we focus on precisely measuring the ratio between the positive kaon decay widths, $R_K = \Gamma (K^+ \to ~e^+\nu)/\Gamma (K^+ \to ~\mu ^+\nu )$, using a stopped kaon beam \cite{cite6}. The E36 experiment is scheduled to obtain physics data at the K1.1BR beam line at the Hadron Experimental Facility of the Japan Proton Accelerator Research Complex (J-PARC) in Tokai, Japan.

We attach special importance to particle identification (PID) in conducting this high precision measurement, which depends on efficiently detecting charged particles (i.e., positrons and positive muons) from kaon decays. The ratio of $K_{e2}$ to $K_{\mu 2}$ events is expected to be approximately 10$^{-5}$. For robust analysis, PID is performed by three independent detectors: time-of-flight (TOF) scintillation counters, threshold-type Cherenkov counters using silica aerogel as a radiator, and lead (Pb) glass Cherenkov counters \cite{cite7}. The use of three independent devices allows us to calibrate the PID capability of each device using the results from the other two. The aerogel Cherenkov (AC) counter was newly designed as a dedicated device for use in the TREK/E36 detector system.

Silica aerogel is an amorphous and porous substance comprised of silica (SiO$_2$) particles and open, air-filled pores on the order of tens of nanometers in size. Recent aerogel production techniques enable to tune the refractive index ($n$) in a wide range from 1.0026 to 1.26 \cite{cite8}. Because of its peculiar, intermediate refractive index and optical transparency, silica aerogel has been widely used as a Cherenkov radiator (see for example Ref. \cite{cite9} as a review). The refractive index of aerogel is approximately related to its density ($\rho $) by $n(\lambda ) - 1 = k(\lambda )\rho $, where $k$ is a constant that depends on the wavelength of light ($\lambda $) \cite{cite10}. In this study, we have developed an aerogel radiator at Chiba University. This aerogel is hydrophobic, and hence requires no maintenance during the experimental period.

\section{Requirements for an E36 aerogel Cherenkov radiator}
\label{}
\begin{figure}[ht]
\centering 
\includegraphics[width=0.45\textwidth,keepaspectratio]{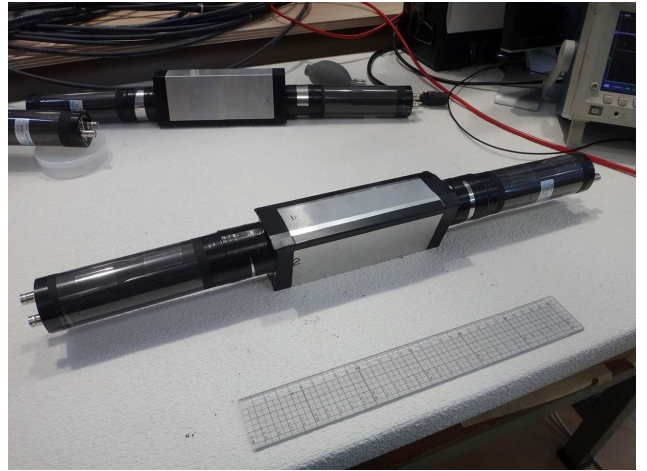}
\caption{Assembled AC counter module. Two PMTs are oriented to the kaon beam axis. The length of the counter housing (aerogel box) is 18 cm.}
\label{fig:fig2}
\end{figure}

\begin{figure}[h]
\centering 
\includegraphics[width=0.40\textwidth,keepaspectratio]{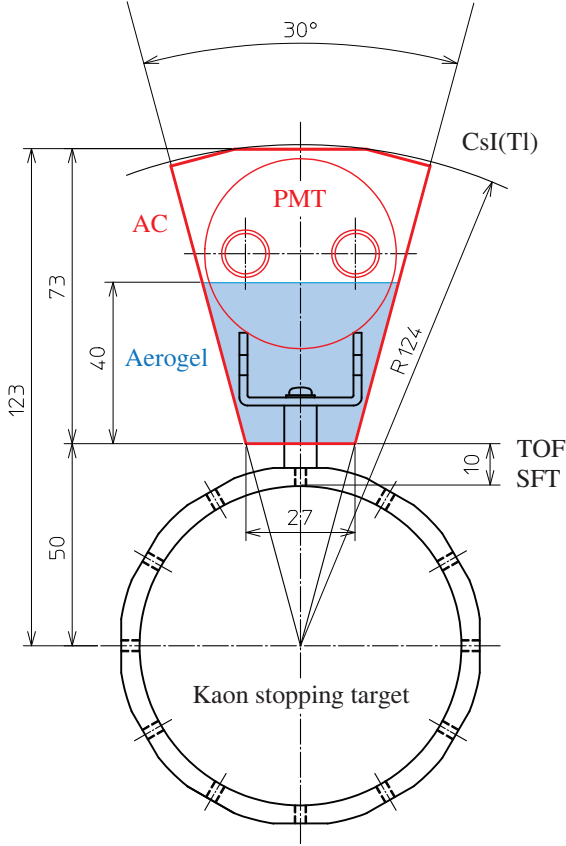}
\caption{A cross-sectional drawing of the central detector perpendicular to the kaon beam axis. The kaon stopping target holder and one AC counter module (red, 12 o'clock direction) with a PMT support are shown. The aerogel radiator block is blue shaded. The spiral fiber tracker (SFT) is directly wound around the target holder. The upstream TOF counter is located over the tracker. The AC counters should be installed between the TOF counter and the inner wall of CsI(Tl) calorimeter.}
\label{fig:fig3}
\end{figure}

The space (i.e., counter height) given to the AC counter is limited to approximately 7 cm between the upstream TOF counters and the CsI(Tl) calorimeter in the central barrel region of the TREK/E36 detector system. To effectively reject events in which the positive muons decay in-flight, we decided to locate the AC counter close to the kaon stopping active target, which is made of plastic scintillating fibers \cite{cite11} and is installed in the central gap of the spectrometer (see Fig. \ref{fig:fig1}). One module of the AC counter is shown in Fig. \ref{fig:fig2}. Fig. \ref{fig:fig3} shows a cross-sectional drawing of the central detector perpendicular to the kaon beam axis. It comprises the target, scintillating-fiber-based spiral fiber tracker \cite{cite12}, upstream TOF counters, and the AC counter. In keeping with the 12 acceptance gaps of the spectrometer, the AC counter is divided into 12 identical modules. Considering the acceptance of the whole detector system, the longitudinal length of the aerogel box of the AC counters is designed to be 180 mm (with an interior length of 179 mm) along the kaon beam axis. Two photomultiplier tubes (PMTs) are attached to both longitudinal sides. To cover the full solid angle around the target, the cross-sectional shape of the aerogel radiator needs to be trapezoidal. Considering a clearance of 0.5 mm at each side, the dimensions of the aerogel blocks to be fabricated were 178 mm in longitudinal length, 46.2 mm in upper-base length, and 24.8 mm in lower-base length, assuming a radiator height (thickness, $t$) of 40 mm.

The refractive index ($n$) of the aerogel radiator needs to be less than 1.095 in order to reject muons with a momentum ($P$) of 236 MeV/$c$ from the $K_{\mu 2}$ decays. In the actual E36 experiment, the $K_{e2}$ ($P_{e^+}$ = 247 MeV/$c$) and $K_{\mu 2}$ events will be accepted by analyzing the charged particle momenta using the spectrometer and charged particle tracking devices, i.e., a spiral fiber tracker and three layers of multiwire proportional chambers. It is desirable to keep a high $n$ value because aerogel radiators with higher $n$ values produce more Cherenkov photons. Conversely, the transparency of the aerogel decreases with increasing $n$ value, resulting in degradation of Cherenkov light collection. Moreover, misidentification of muons as positrons due to Cherenkov radiation caused by knock-on electrons ($\delta $-rays) may increase in a dense aerogel. Our requirements for the AC counter include a positron detection efficiency greater than 98\% and a positive muon misidentification rate lower than 3\%. During 2010--2013, in search for the best solution for aerogel specification (e.g., $n$ = 1.037, 1.05, and 1.08) as well as the best counter configuration (e.g., specular/diffusive reflective sheets of inner wall and mirror shape on the inside of the outer wall), we performed a series of test beam experiments using prototype counter modules at the Research Center for Electron Photon Science at Tohoku University in Japan, the National Laboratory for Particle and Nuclear Physics (TRIUMF) in Canada, and J-PARC. From the results of the test experiments \cite{cite1,cite13,cite14}, the design of the AC counter module was finalized to use a specular reflective sheet (aluminized Mylar), and the specification of the aerogel radiator was proposed to be $n$ = 1.08 and $t$ = 40 mm. A deviation from $n$ = 1.08 will not have a significant impact on the detector performance; e.g., the final spread of $\pm $0.004 (5\%) in the produced refractive index for each counter is acceptable.

\section{Aerogel fabrication}
\label{}
Our method for producing the silica aerogel blocks to be used in the E36 experiment was based on a modified conventional technique described in Ref. \cite{cite10}. First, a wet gel was synthesized by means of the sol--gel method in an appropriate mold described in Section 3.1. To obtain highly transparent aerogel, the classic KEK method \cite{cite15} (which uses ethanol or methanol as a solvent) was modified by introducing the solvent $N$,$N$-dimethylformamide (DMF) into the wet-gel synthesis step \cite{cite16}. To attain aerogel with $n \sim $ 1.08, we used only DMF as the solvent, whereas we generally used a mixture of DMF and methanol at $n \sim $  1.05. After aging the wet gel, it was detached from the mold in an ethanol bath, and we performed a hydrophobic treatment by adding hexamethyldisilazane into the ethanol bath \cite{cite17}. After removing impurities from the wet gel by repeatedly replacing the ethanol, we finally obtained an aerogel using the supercritical carbon dioxide drying method.

\subsection{Molding}

\begin{figure}[t]
\centering 
\includegraphics[width=0.45\textwidth,keepaspectratio]{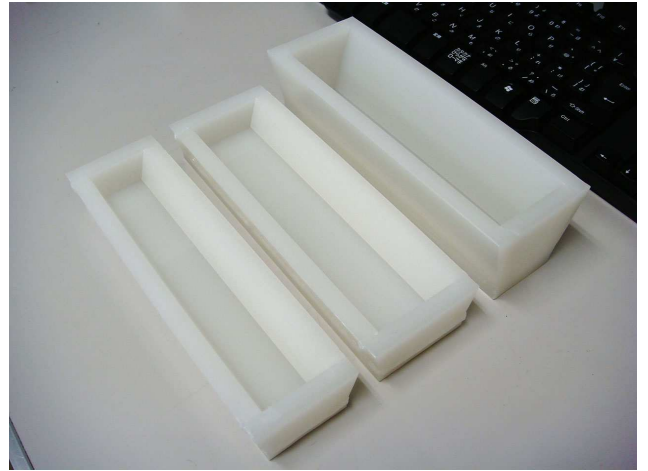}
\caption{Custom-made molds for the (left) upstream layer and (center) downstream layers made of polypropylene. Both molds measure approximately 182 mm in longitudinal inner length and 21.5 mm in inner depth. Prior to the final production, the right mold was used for a test production of 4 cm-thick blocks (see Appendix A).}
\label{fig:fig4}
\end{figure}

\begin{figure}[t]
\centering 
\includegraphics[width=0.45\textwidth,keepaspectratio]{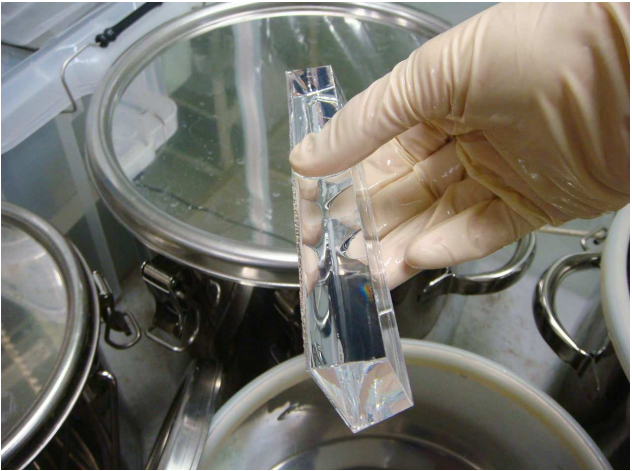}
\caption{A wet gel bar detached from the mold. When the aerogel density was more than $\sim $0.1 g/cm$^3$ ($n$ = 1.08 corresponding to $\rho $ = 0.27 g/cm$^3$), direct handling of the wet gel was possible. The wet gel needed to be returned to the ethanol bath within several tens of seconds to avoid cracking because of drying.}
\label{fig:fig5}
\end{figure}

Aerogel blocks with a trapezoidal cross-section can be produced in two ways: cutting or molding. The cutting method uses a water jet cutter on square aerogel tiles with dimensions of approximately 11 $\times $ 11 $\times $ 2 cm$^3$ to produce trapezoidal prism aerogel segments, making full use of the hydrophobic feature of our aerogel. In this case, one block comprises a minimum of 4 trapezoidal prism segments. Commercially available square molds are economical. Conversely, the molding method uses a trapezoidal mold for synthesizing a wet gel. In the latter case, to suppress cracking of the aerogel blocks during supercritical drying, the thickness of one aerogel bar should be reduced to 2 cm by dividing one trapezoidal aerogel block with a thickness of 4 cm into two layers (see Appendix A). This method requires custom-made molds. The cost for manufacturing the custom-made molds is almost equivalent to that of machining the aerogel tiles with water jets.

Considering the total cost, including manpower and time, molding is more efficient for a large production. When we manufactured aerogel in-house, the supercritical drying was the most manpower-intensive and time-consuming process for which we used our supercritical carbon dioxide drying apparatus with a 7.6 l autoclave. Aerogel cracking often appeared during this process. We assumed that the crack-free yield of aerogel blocks with $n \sim $ 1.08 would be 70\%, independent of the wet gel shape. The cutting and molding methods require four and three supercritical drying operations, respectively, to obtain the necessary aerogel blocks as well as several spares. More specifically, for 12 whole blocks and 2 spares, we had to produce 24 aerogel tiles without cracking (14 tiles for the larger downstream segments and 10 tiles for the smaller upstream segments near the kaon stopping target) by cutting, or 28 bars by molding. Considering the crack-free yield, we had to synthesize 35 wet gel tiles or 40 wet gel bars. Our autoclave for supercritical drying could store 10 wet gel tiles or 14 wet gel bars at a time.

Molding had the important advantage of enabling us to obtain clear aerogel surfaces. Under cutting, the water jet-machined aerogel surface became significantly rough and scattered laser beams so that it was no longer possible to measure the refractive index. The aerogel surface could not be polished. Molding allowed us to create a very clear aerogel surface by keeping the inner surface of the mold flat. Our counter works best if all aerogel surfaces are clear, and requires crack-free aerogel blocks because it was designed to be lined with aluminized Mylar to induce specular reflection of Cherenkov light at the inside wall.

Another advantage of molding is that an entire trapezoidal block could be made from only two aerogel parts. Under cutting, 4 segments (i.e., upstream and downstream layers, each segmented to two parts) were needed to form a whole trapezoidal prism block for one module. That was because the dimensions of the synthesized square wet gel were limited by the size of the commercially available square mold and our autoclave. Under molding, one trapezoidal block module could be made from two-layer semi-monolithic aerogel bars. We placed the long wet gel bars vertically in the autoclave because the autoclave was of sufficient depth (30 cm) (see Appendix A). Semi-monolithic aerogel blocks facilitate fixing them to the counter housing. To secure an air light guide gap between the aerogel blocks and counter roof (see Fig. \ref{fig:fig3}), the radiator is held to the bottom of the housing using fixtures. Because we can fix the radiator with a small number of fixtures, monolithic aerogel block is important, especially for the counter module in six o'clock direction, where it is arranged roof-side down around the kaon stopping target.

Considering also the results of the pilot production described in Appendix A, we finally opted for the molding technique for producing aerogel blocks for use in the actual counters. To produce trapezoidal prism aerogel bars, we devised custom-made molds. The mold was an open-topped box into which we could pour the prepared chemical solution. It was manufactured by welding smooth polypropylene plates with a thickness of 10 mm using welding bars by Tokiwa Co., Ltd., Japan. To form the whole trapezoidal block from two-layer aerogel bars, we prepared ten copies each of two different sizes of mold; one for the upstream (small) and one for the downstream (large) layers (Fig. \ref{fig:fig4}). Both molds had the same length, but different widths of their trapezoidal sides. We expected the longitudinal shrinkage ratio of the aerogel bars to be 0.975 in the production process and designed the dimensions of the mold accordingly. The molds were manufactured with a dimensional accuracy greater than 1 mm.

\subsection{Chemical preparation recipe}

\begin{table*}[ht]
\centering 
\caption{Chemical solutions used in wet gel synthesis for each aerogel bar.}
\label{table:table1}
	\begin{tabular}{ll}
		\hline
		Chemicals & Dose [g] for upstream (downstream) layer \\
		\hline
		Polymethoxy siloxane$^a$ & 39.05 (52.85) \\
		Distilled water & 22.52 (30.48) \\
		$N$,$N$-Dimethylformamide & 57.23 (77.46) \\
		28\% Ammonia solution & 0.24 (0.33) \\
		\hline
		\multicolumn{2}{l} {$^a$Methyl silicate 51 (Fuso Chemical Co., Ltd., Japan).} \\
	\end{tabular}
\end{table*}

Table \ref{table:table1} lists the preparation recipe for the raw chemicals for producing aerogel bars with $n$ = 1.08. This recipe allowed us to obtain samples with $n$ = 1.076 in the experimental production of square tiles (see Appendix A). The use of DMF as the solvent was helpful in attaining highly transparent aerogel with a high refractive index. Based on the performance of prototype aerogel samples as Cherenkov radiators measured with test beams \cite{cite14}, we chose the recipe for fabricating the aerogel blocks for the actual detector. Wet gel slightly shrank in the aging process followed by their synthesis, and depending on the refractive index (i.e., the density of the silica matrix), it also shrank during the supercritical drying process \cite{cite10}. The refractive index of the final products slightly depended on their volume and shape. Apart from the above shrinkage factor, the refractive index of the aerogel blocks was basically determined by the preparation recipe of the raw chemicals. Namely, the density of aerogel depends on the volume ratio of the solvent used in the wet gel synthesis.

\subsection{Final production}

\begin{figure}[t]
\centering 
\includegraphics[width=0.45\textwidth,keepaspectratio]{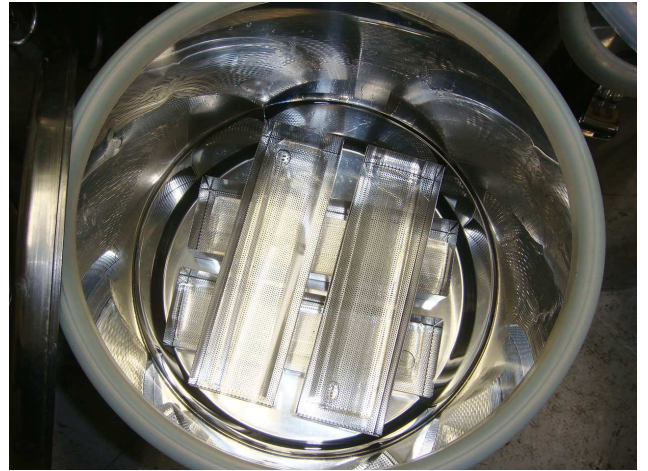}
\caption{Wet gel in the hydrophobic treatment process. The wet gel bars were placed in the punched trays and soaked in a solution for the hydrophobic treatment.}
\label{fig:fig6}
\end{figure}

\begin{figure}[t]
\centering 
\includegraphics[width=0.45\textwidth,keepaspectratio]{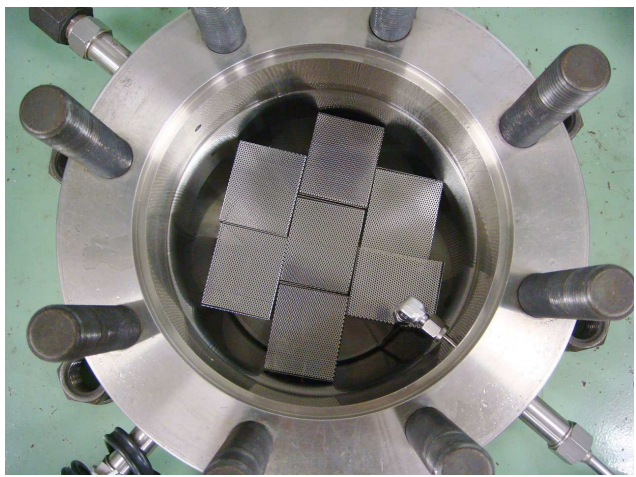}
\caption{Punched trays placed in the autoclave of supercritical carbon dioxide drying apparatus. Each tray contained two wet gel bars. The autoclave was filled with ethanol at the beginning of the drying operation.}
\label{fig:fig7}
\end{figure}

From April to June 2014, we produced aerogel bars with $n$ = 1.08 for use in the actual detector. Dividing the whole production into three lots, 25 and 20 wet gel bars for the upstream and downstream layers were synthesized, respectively. The molds were reused a maximum of three times by cleaning them after each use. Immediately after mixing the chemical solutions shown in Table \ref{table:table1} in a beaker, the prepared solution was strongly stirred for 30 s, immediately poured into a mold, and covered by a lid. From the specific gravity of each chemical, the volumes of the solution were calculated to be 116 and 157 ml for the upstream and downstream layers, respectively, corresponding to a 20.5 mm wet gel thickness. At room temperature (22--26$^\circ $C), we predetermined the amount of ammonia solution as a catalyst, so that the solution gelled approximately two minutes after the beginning of mixing. After a further two minutes, the surface of the wet gel synthesized in the mold was filled with 4--6 ml of methanol to prevent it from drying; it was then covered with a 0.3 mm-thick aluminum plate and aged in a sealed tank for one week.

The detachment of wet gel from the molds was the key process in wet gel molding. To facilitate the detachment, the wet gel in the mold was aged in the sealed tank filled with ethanol for an additional day. Soaking the wet gel in ethanol promoted its shrinkage. Moreover, by soaking the mold into the ethanol upside down, the wet gel was spontaneously detached from the mold because of its own weight (Fig. \ref{fig:fig5}).

\begin{figure}[ht]
\centering 
\includegraphics[width=0.45\textwidth,keepaspectratio]{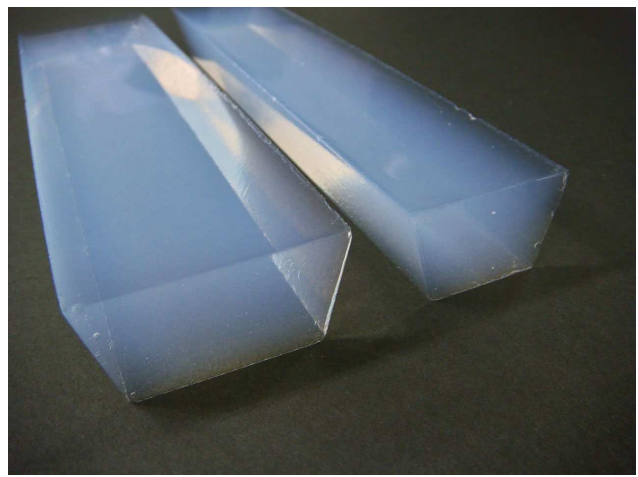}
\caption{Crack-free aerogel bars for (left) downstream and (right) upstream layers obtained for the final production. Both the aerogel bars had a longitudinal length of approximately 18 cm and a thickness of 2 cm.}
\label{fig:fig8}
\end{figure}

After removing the mold, the wet gel was subjected to the hydrophobic treatment. The wet gel was transferred into a stainless steel punched tray specially manufactured for this study and soaked temporarily in another ethanol bath to prevent it from cracking due to drying. By adding hexamethyldisilazane into the ethanol bath (which was used for additional aging of the wet gel) and stirring, the solution for the hydrophobic treatment was prepared, with the volume ratio of the hydrophobic reagent to ethanol being approximately 1:9 \cite{cite10}. The wet gel was soaked in the solution for the hydrophobic treatment for 3--4 days, as shown in Fig. \ref{fig:fig6}. To reduce impurities other than ethanol in the wet gel, the hydrophobic reagent/ethanol filling the tank was replaced three times with new ethanol.

Three operations of the supercritical carbon dioxide drying apparatus yielded 42 aerogel bars. The autoclave in the apparatus was filled with new ethanol, and the wet gel bars were placed there by standing them vertically on the punched trays, as shown in Fig. \ref{fig:fig7}. The punched trays were designed so that seven of them could be installed in the autoclave. Two wet gel bars (basically a combination of the upstream and downstream bars) could be placed in each tray, i.e., 14 wet gel bars could be dried at once. The operation process was based on Ref. \cite{cite10}, where the rate of temperature rise from 40 to 80$^\circ $C was reduced from 10$^\circ $C/h to 5$^\circ $C/h to suppress cracking of the aerogel bars. We emphasize that a slow pressure reduction rate (below 1 MPa/h) was also adopted in the end process of the operation for the same purpose.

\begin{figure}[ht]
\centering 
\includegraphics[width=0.45\textwidth,keepaspectratio]{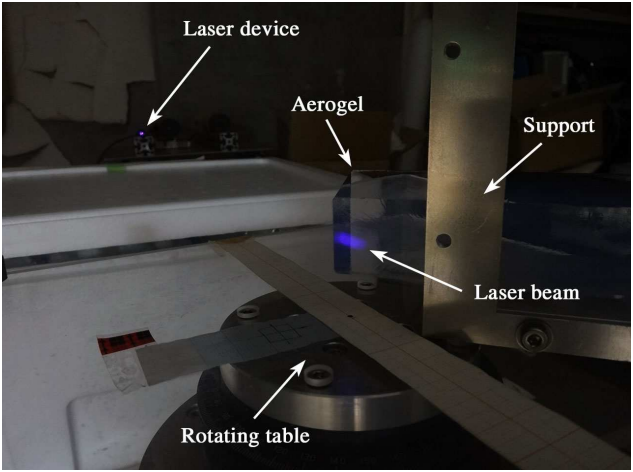}
\caption{Setup for refractive index measurement. An aerogel bar was arranged in the laser beam line on the rotating table using the aluminum support. The minimum deviation of the laser beam was measured on a screen approximately 1.8 m downstream of the rotating table.}
\label{fig:fig9}
\end{figure}

\section{Optical characterization}
\label{}

\subsection{Cracking}
We obtained 30 crack-free aerogel bars out of the 42 bars produced after the three supercritical drying operations. More specifically, 16 out of 24 aerogel bars and 14 out of 18 bars for the upstream (small) and downstream (large) layers, respectively, had no cracking. We ensured the required number of aerogel blocks for the 12 counter modules and the two spare sets. The obtained aerogel bars had an impressive appearance, as shown in Fig. \ref{fig:fig8}. The number of aerogel bars with cracking from the first, second, and third supercritical-drying batches (not synthesis lots) were 5, 5, and 2 bars out of 14 bars per batch. This slight batch dependence of cracking can be due to the manual operation (i.e., manual pressure control) of the drying apparatus. Cracking did not depend significantly on the aerogel size.

\begin{figure}[ht]
\centering 
\includegraphics[width=0.50\textwidth,keepaspectratio]{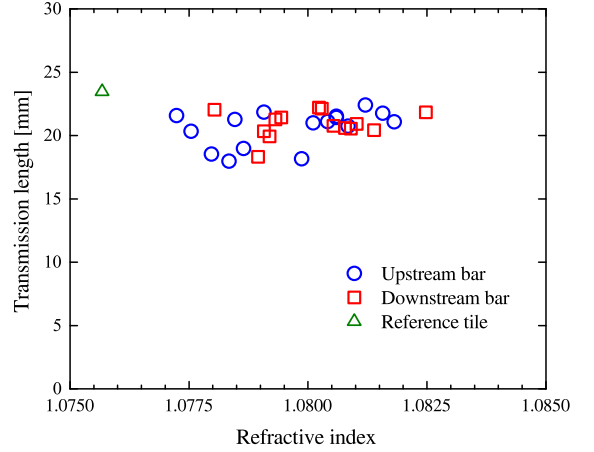}
\caption{Distribution of transmission length at $\lambda $ = 400 nm (see Section 4.3) as a function of the refractive index. The refractive index was measured at $\lambda $ = 405 nm. A total of 30 aerogel bars for upstream and downstream layers are represented by circles and squares, respectively. For reference, the square aerogel tile is shown by a triangle.}
\label{fig:fig10}
\end{figure}

\subsection{Refractive index}
We measured the refractive index of the final production of aerogel bars using the laser Fraunhofer method described in Ref. \cite{cite10}. The method allowed to determine the refractive index at the corner of the aerogel blocks by measuring the deviation of the laser path. A blue--violet semiconductor laser with $\lambda $ = 405 nm was used. When measuring the refractive index of square aerogel tiles, each corner of the tiles was generally irradiated with the laser beam. For the trapezoidal prism aerogel bars produced, we exposed the right angle between the bottom surface and the trapezoidal side surface to the laser beam, as shown in Fig. \ref{fig:fig9}. The minimum distance between the laser path in the aerogel and the edge, defined as the side between the bottom surface and the trapezoidal side surface of the aerogel, was set to be 5 mm. The refractive indices measured at both ends of an aerogel bar were then averaged.

We successfully obtained aerogel bars with the desired refractive index. The measured refractive index was distributed in a range between 1.0772 and 1.0825 for the 30 crack-free aerogel bars (Fig. \ref{fig:fig10}). As a reference, we fabricated a square tile with dimensions of approximately 9 $\times $ 9 $\times $ 2 cm$^3$ at the same time as the first lot. The refractive index of this reference aerogel was 1.0757, which was smaller than that of the trapezoidal prism aerogel. This suggests that the measured refractive index depended partially on the volume and the shape of aerogel blocks, where the synthesis volumes of the upstream (downstream) trapezoidal wet gel bars and the square one were approximately 116 (157) and 187 ml, respectively. In general, the refractive index was inversely proportional to the synthesis volume of the wet gel blocks. The macroscopic shape and size of wet gel could have an influence on the nanostructure formation and wet gel shrinkage in the synthesis and aging processes.

\begin{figure}[th]
\centering 
\includegraphics[width=0.45\textwidth,keepaspectratio]{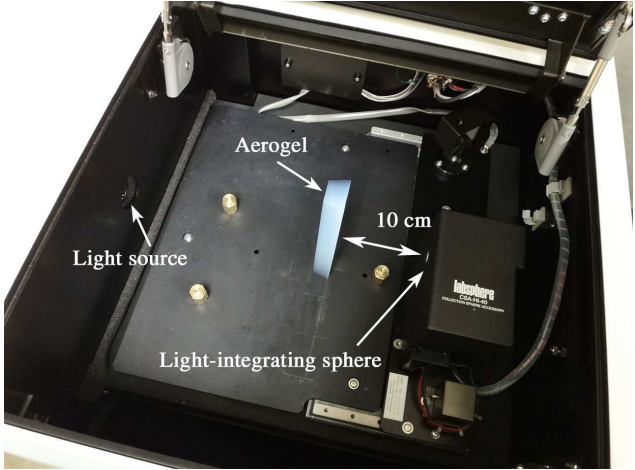}
\caption{Measurement setup in the light-shielded chamber of the Hitachi U-4100 spectrophotometer. Light transmission along the aerogel thickness direction was measured. The bottom surface of the aerogel (see Fig. \ref{fig:fig8}) corresponding to the upstream side in the AC counter was placed at the upstream side of the spectrophotometer. The distance between the aerogel's downstream surface and the entrance of a light-integrating sphere was set to be 10 cm.}
\label{fig:fig11}
\end{figure}

\subsection{Transparency}
\begin{figure}[hb]
\centering 
\includegraphics[width=0.50\textwidth,keepaspectratio]{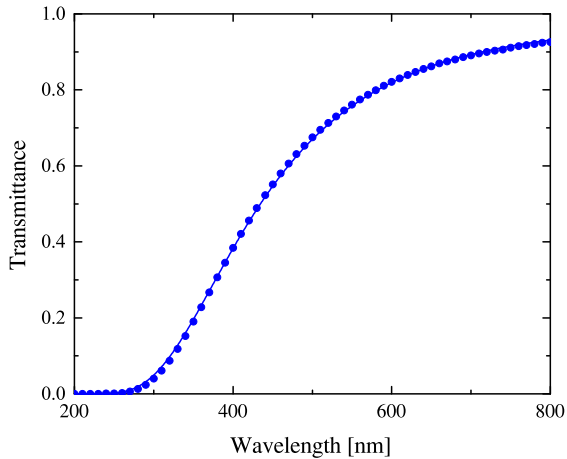}
\caption{Transmittance curve for a typical aerogel ($t$ = 19.9 mm) as a function of wavelength. Circles show the transmittance measured every 10 nm, and the solid curve shows the fit with $T=A\exp(-Ct/\lambda ^4)$. The parameters obtained from the fitting were $A$ = 0.988 $\pm $ 0.001 and $C$ = 0.01220 $\pm $ 0.00005 $\mu $m$^4$/cm.}
\label{fig:fig12}
\end{figure}

We measured the transmittance of the produced aerogel bars at wavelengths ranging from 200 to 800 nm using a spectrophotometer U-4100 (Hitachi, Ltd., Japan). Fig. \ref{fig:fig11} shows the measurement setup in the light-shielded chamber of the spectrophotometer. To detect as little of the light scattered in the aerogel as possible, the distance between the aerogel's downstream surface and the entrance of a light-integrating sphere was set to be 10 cm \cite{cite10}. Fig. \ref{fig:fig12} shows the measured transmittance of a typical aerogel bar from the final production as a function of wavelength. The mean transmittance at $\lambda $ = 400 nm, through a thickness of approximately 20 mm, was 38.4\% for the 30 crack-free aerogel bars. In the case in which we arranged a combination of upstream and downstream aerogel bars with a total thickness of approximately 40 mm, the transmittance was 13.1\%.

Light transmission in aerogel is known to be dominated by Rayleigh scattering:
\[ T(\lambda , t)=A\exp(-Ct/\lambda ^4), \]
where $T$ is the transmittance, $A$ is the amplitude, and $C$ is called the ``clarity coefficient,'' and it is usually measured in units of $\mu $m$^4$/cm. The clarity coefficient obtained from the fitting was $C$ = 0.01220 $\pm $ 0.00005 $\mu $m$^4$/cm for the above typical aerogel sample (Fig. \ref{fig:fig12}).

The calculated transmission length, defined as $\Lambda _{\rm T}(\lambda ) = -t/{\rm ln}T(\lambda )$ of the aerogel bars at $\lambda $ = 400 nm was reasonable, considering the refractive index ($n \sim $ 1.08). The mean transmission length of the 30 aerogel bars at $\lambda $ = 400 nm was 20.8 mm (see Fig. \ref{fig:fig10}). This value is consistent with that plotted in the transmission length--refractive index scatter graph shown in Ref. \cite{cite10}. The reference square aerogel tile had $\Lambda _{\rm T}$ = 23.5 mm.

\subsection{Dimensions}

\begin{figure}[t]
\centering 
\includegraphics[width=0.45\textwidth,keepaspectratio]{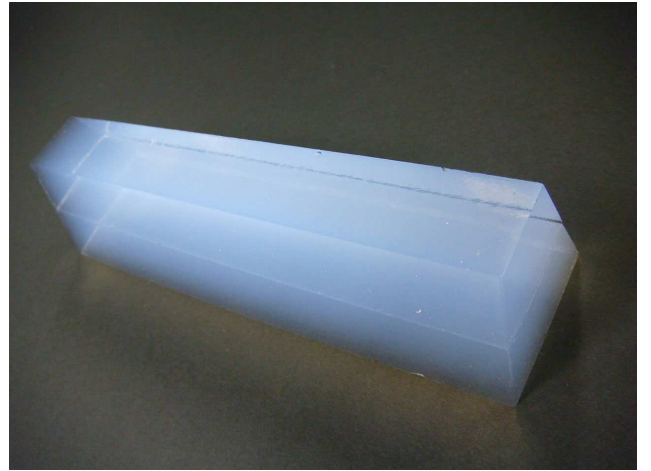}
\caption{Upstream aerogel bar stacked on top of a downstream one. The longitudinal length and total thickness are approximately 18 and 4 cm, respectively.}
\label{fig:fig13}
\end{figure}

The dimensions of the produced aerogel bars met our requirements. The longitudinal length of the bars ranged from 176.3 to 177.8 mm for the 30 crack-free aerogel bars, which was consistent with our expectation (i.e., 178 mm). The lower- (upper-) base length of the cross-sectional trapezoid of the upstream (downstream) layer was 24.0--24.5 (45.5--46.4) mm, which is in good agreement with the requirements (i.e., 24.8 (46.2) mm). As shown in Fig. \ref{fig:fig13}, the upstream aerogel bar was flush with the downstream one; thus, they will form a radiator unit in the counter module.

\begin{figure}[t]
\centering 
\includegraphics[width=0.50\textwidth,keepaspectratio]{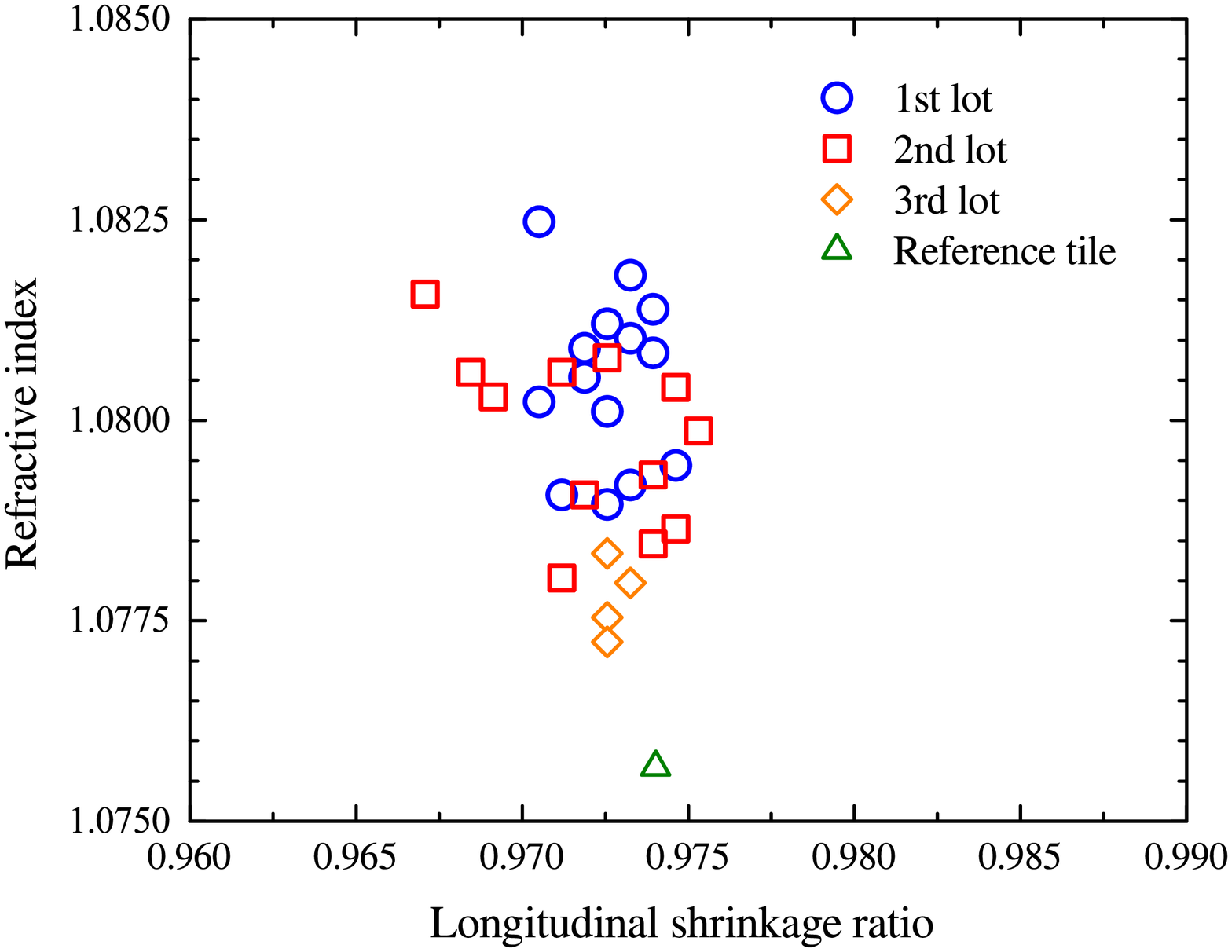}
\caption{Refractive index measured at $\lambda $ = 405 nm as a function of longitudinal shrinkage ratio for each of the 30 crack-free aerogel bars and one reference square tile. The aerogel bars are classified based on their wet-gel synthesis lot, indicated by different symbols.}
\label{fig:fig14}
\end{figure}

The mean longitudinal shrinkage ratio was 0.972, close to our expectation of 0.975, where the longitudinal length of the mold was taken to be 182.25 mm based on the actual measurement. Fig. \ref{fig:fig14} shows the refractive index as a function of the longitudinal shrinkage ratio. There is a tendency for the refractive index to increase with decreasing longitudinal shrinkage ratio. In addition, the refractive index depends on the wet gel lot in which it was synthesized, especially between the first and third lots. This could be due to the difference in room temperature during the production process. The third lot was fabricated in a slightly high-temperature environment (24--26$^\circ $C) compared with the first lot (21--24$^\circ $C).

\section{Conclusion}
\label{}
We have developed hydrophobic silica aerogel with $n$ = 1.08 to be used as a radiator in threshold-type Cherenkov counters. These counters are meant to separate the positrons from positive muons with a momentum of approximately 240 MeV/$c$ produced by kaon decays in the J-PARC TREK/E36 experiment. The requirements for the Cherenkov radiator were determined by the results of test beam experiments and the design of the counter configuration. We have described a method for producing aerogel bars with a trapezoidal cross-section and a length of 18 cm to fit the barrel region surrounding the kaon stopping target of the TREK/E36 detector system. Production of the aerogel bars for the actual detector made up of 12 counter modules was successfully performed by dividing each radiator volume into two layers with a total thickness of 4 cm. The block dimensions and optical parameters, including a transmission length at 400 nm wavelength of approximately 20 mm, have been measured and found suitable for use in the actual detector.

\section*{Acknowledgments}
\label{}
The authors are grateful to the members of the J-PARC TREK/E36 Collaboration for fruitful discussions on aerogel development. We are also grateful to Dr. H. Nanjo of Kyoto University for his assistance in designing the aerogel mold. We performed the optical measurements of aerogel at KEK; we are thankful to Prof. I. Adachi for his support. We are also thankful to the Venture Business Laboratory at Chiba University for offering room to manufacture the aerogel. This work was supported by a Grant-in-Aid for Scientific Research (B) (No. 25287064) from the Japan Society for the Promotion of Science (JSPS). M. Tabata was supported in part by the Space Plasma Laboratory at the Institute of Space and Astronautical Science (ISAS), Japan Aerospace Exploration Agency (JAXA).

\appendix

\section{Experimental fabrication of aerogel with $n$ = 1.08}
\label{}
In April--June, 2012, we tested if we could dry a long aerogel bar with no cracking by placing it vertically into the autoclave of a supercritical carbon dioxide drying apparatus. Autoclaves have a vertically long depth to reduce the aperture area while keeping the total volume. In general, we stack wet gel tiles with a typical thickness of 2 cm from the bottom of the autoclave. Vertically arranging long wet gel bars to make full use of the autoclave's depth was our first challenge. We synthesized wet gel bars 2 cm in diameter and 15 cm in length using a chemical recipe for $n \sim $ 1.08 aerogel and acrylic hollow pipes as mold. As a result of drying, the experimental production of 15 cm long aerogel bars was successful.

The chemical preparation recipe for aerogel with $n$ = 1.08, shown in Table \ref{table:table1}, was determined by comparing two slightly different recipes. These recipes were basically estimated by interpolating the recipes used to produce the aerogel samples with $n$ = 1.045--1.11, presented in Ref. \cite{cite10}. In September, 2012, we experimentally fabricated eight square aerogel tiles with dimensions of approximately 11 $\times $ 11 $\times $ 2 cm$^3$ with the recipes targeting an index of $n$ = 1.08. One recipe had a 3\% higher silica precursor (polymethoxy siloxane) content compared with the other recipe, resulting in aerogel tiles with $n$ = 1.076 and 1.074, respectively. Thus, we chose the former recipe for the production.

Prior to the final production, an item to be considered was the thickness of trapezoidal prism aerogel bars produced under a molding method. In February--April, 2014, we performed a series of pilot productions of various aerogel tiles and bars with $n$ = 1.08 to finalize the details of the fabrication method. To fill one counter module with a monolithic aerogel block, seven trapezoidal wet gel bars 18 cm in length and 4 cm in thickness were synthesized using a special mold manufactured by the same method described in Section 3.1 (also see Fig. \ref{fig:fig4}); however, we failed to obtain good aerogel blocks because of cracks caused during the fabrication processes (e.g., hydrophobic treatment or supercritical carbon dioxide drying). Therefore, we had to separate one module into two layers with thicknesses of $t$ = 2 cm under the molding method. In fact, in the next trial fabrication, we succeeded in producing the first crack-free aerogel bar with a reduced thickness of 2 cm for the upstream layer using the same mold. Thus far, it is difficult to produce square aerogel tiles with thicknesses exceeding 2 cm without cracking caused by the supercritical carbon dioxide drying.

The other item to be considered was the use of the supercritical ``ethanol'' drying method. So far, we have usually used this method to fabricate low-density aerogel \cite{cite18}. Our apparatus for the supercritical ethanol drying uses an autoclave larger (30 l) than that of the carbon dioxide drying apparatus and semi-automated pressure control valves. Supercritical ethanol drying provides cost-effective production, as well as a shorter operation time than carbon dioxide drying. Motivated by the recent successful production of large-area aerogel tiles with up to $n$ = 1.06 using the ethanol method, we dried three square wet gel tiles with $t$ = 2 cm and one wet gel bar with $t$ = 4 cm by the supercritical ethanol method. Only low-quality aerogel was obtained, i.e., all samples were broken due to major cracks and/or warpage. We concluded that we must use the supercritical carbon dioxide drying method to fabricate aerogel with $n$ = 1.08.




\bibliographystyle{model1-num-names}

\begin{thebibliography}{00}
\bibitem{cite1}
J-PARC TREK Collaboration, J-PARC E36 Experimental Proposal, 2010, and Addenda, 2011 and 2012, http://trek.kek.jp/e36/publication.html
\bibitem{cite2}
C. Djalali (TREK Collaboration), AIP Conf. Proc. 1423 (2012) 297.
\bibitem{cite3}
J.A. Macdonald, et al., Nucl. Instrum. Methods A 506 (2003) 60.
\bibitem{cite4}
M. Abe, et al. (KEK-E246 Collaboration), Phys. Rev. D 73 (2006) 072005.
\bibitem{cite5}
A. Kawachi, et al., Nucl. Instrum. Methods A 416 (1998) 253. 
\bibitem{cite6}
S. Shimizu (TREK Collaboration), AIP Conf. Proc. 1441 (2012) 338.
\bibitem{cite7}
Y. Miyazaki, et al., Nucl. Instrum. Methods A 779 (2015) 13.
\bibitem{cite8}
M. Tabata, et al., Nucl. Instrum. Methods A 623 (2010) 339.
\bibitem{cite9}
C. Lippmann, Nucl. Instrum. Methods A 666 (2012) 148.
\bibitem{cite10}
M. Tabata, et al., Nucl. Instrum. Methods A 668 (2012) 64.
\bibitem{cite11}
Y. Igarashi (TREK Collaboration), PoS(KAON09)015.
\bibitem{cite12}
M. Tabata, et al., JPS Conf. Proc. (2015) in press.
\bibitem{cite13}
A. Toyoda, et al., ELPH Annual Report, Tohoku University 1 (2010) 101.
\bibitem{cite14}
A. Toyoda, et al., ``Performance evaluation of aerogel Cherenkov detector for the J-PARC E36 experiment,'' presented at the Physical Society of Japan 2013 Autumn Meeting, 21aSL-9, Kochi, Japan, September 2013 [in Japanese], http://kds.kek.jp/getFile.py/access?contribId=79\&sessionId=28\&resId=\\0\&materialId=slides\&confId=13773
\bibitem{cite15}
I. Adachi, et al., Nucl. Instrum. Methods A 355 (1995) 390.
\bibitem{cite16}
I. Adachi, et al., Nucl. Instrum. Methods A 553 (2005) 146.
\bibitem{cite17}
H. Yokogawa and M. Yokoyama, J. Non-Cryst. Solids 186 (1995) 23.
\bibitem{cite18}
M. Tabata, et al., Biol. Sci. Space 25 (2011) 7.
\end{thebibliography}



\end{document}